\documentclass{jpsj-suppl}
\usepackage{txfonts} 

\title{Pairing in Nuclei: Exact Solutions}

\author{A.B. \textsc{Balantekin}$^{1}$}

\inst{$^{1}$University of Wisconsin, Department of Physics, Madison, Wisconsin 53706 USA}

\email{baha@physics.wisc.edu}

\recdate{March 31, 2018}

\abst{Pairing plays an essential role in describing nuclear spectra and attempts to describe it has a long history in nuclear physics. Many theoretical tools were developed to treat the pairing problem either exactly or at various levels of approximation. In this contribution to the proceedings of a conference honoring the career of Taka Otsuka,  recent developments in the exact solutions of the pairing problem are described.}

\kword{Pairing in many-body systems, nuclear pairing, collective neutrino oscillations, Bethe ansatz}

\begin{document}
\maketitle

\section{Introduction}

Pairing of fermions is a common feature of many-body systems. Studies of pairing of like nucleons have a long history in nuclear physics 
\cite{Dean:2002zx}. Since the advent of facilities with rare isotope beams, it is now possible to probe not only neutron-neutron and proton-proton, but also proton-neutron pairing. As we describe below, the pairing behavior is present in systems as varied as nuclei, condensed-matter systems, and many-neutrino gases.  

A one-body term in the Hamiltonian consisting of a fermion and another one in a suitably-defined conjugate state can be the starting point of a generalized definition of pairing. For example in nuclear physics L=0 (s-wave) pairing 
incorporates a nucleon and another one in a time-reversal state ($|j,m \rangle$ vs. $|j,-m \rangle$) \cite{Otsuka:1978zz}. In a more elaborate picture one can introduce generalized seniority, L=2 (d-wave) pairing, etc. Indeed s- and d-wave pairing of like nucleons describe the collectivity  of low-lying states of many medium-heavy even-even nuclei \cite{Iachello:1979jaq}. 
A pairing interaction is a two-body term in the Hamiltonian which contains two pairs (or mathematical equivalent). Pairing Hamiltonians may  display many symmetries which help finding eigenvalues and eigenstates of such Hamiltonians. 

Perhaps the most ubiquitous example of pairing is the BCS theory of superconductivity, which is widely used both in condensed-matter and nuclear physics.  Fermion mass term in the Dirac Hamiltonian is similar to L=0 pairing in nuclear physics except it includes parity conjugate fields (left-handed and right-handed components of the Dirac spinor). For charged fermions this term takes the form $a^{\dagger}_L a_R + h.c.$, the Dirac mass term. 
For neutral fermions charge conservation is not an issue, so one can have a Majorana mass, $a^{\dagger}_L a^{\dagger}_R + h.c.$ Combining Dirac and Majorana mass terms  (like combining particle-particle and hole-hole pairing with particle-hole pairs in nuclear physics) leads to the SO(5) symmetry 
\cite{Balantekin:2000qt}. This symmetry results in seesaw mechanism for neutrino masses. 
Neutrino oscillations include family-symmetry conjugate pairs (e.g. electron neutrino and muon neutrino). Associated SU(2) algebra is the neutrino flavor isospin ($\nu_e$ is "spin-up", $\nu_{\mu}$ is "spin-down"). Majorana type pairing has not yet been observed through double beta decay, but many examples of that type interaction was recently uncovered in condensed matter physics. In astronomical sites with a large number of neutrinos (such as core-collapse supernovae and neutron star mergers) a pairing type interaction (representing forward scattering of neutrinos off other neutrinos) governs neutrino transport 
with significant implications in nucleosynthesis of elements  \cite{Balantekin:2006tg}. These interactions alter the $\nu_e$ and $\overline{\nu}_e$ content of the neutrino flux and consequently $n/p$ ratio through the reaction 
$\nu_e + n \rightarrow p + e^-$ and the reaction $\overline{\nu}_e + p \rightarrow n + e^+$.

\section{Pairing and Quasi-spin Algebra}

\subsection{Quasispin algebra} 
The quasi-spin algebra is defined in terms of fermion creation and annihilation operators as 
\begin{eqnarray}
\hat{S}^+_j&=&\sum_{m>0} (-1)^{(j-m)} a^\dagger_{j\>m}a^\dagger_{j\>-m},
\nonumber \\
\hat{S}^-_j&=&\sum_{m>0} (-1)^{(j-m)} a_{j\>-m}a_{j\>m}, \nonumber
\end{eqnarray}
\begin{equation}
\hat{S}^0_j=\frac{1}{2}\sum_{m>0}
\left(a^\dagger_{j\>m}a_{j\>m}+a^\dagger_{j\>-m}a_{j\>-m}-1
\right) .
\end{equation}
In these equations $j$ is the quantum number of the Shell Model orbital. 
This quasispin algebra  is an SU(2) algebra with quadratic Casimir operator $\Omega_j (\Omega_j/2+1)/2$, where $\Omega_j = j+1/2$ is the maximum number of pairs that can occupy the level $j$. 

\subsection{Exact Solutions}

One can write the Hamiltonian describing a general pairing problem as 
\begin{equation}
\label{general}
\hat{H}=\sum_{jm} \epsilon_j a^\dagger_{j\>m} a_{j\>m} -
\sum_{jj'} G_{j j'} \hat{S}^+_j \hat{S}^-_{j'} ,
\end{equation}
where $\epsilon_j$ and $G_{j j'}$ represent single particle energies and the pairing interaction strength, respectively. 
The pairing Hamiltonian of Eq. (\ref{general}) is not easily solvable, except for a few special cases. The quasispin algebra was originally introduced to solve the limit in which one can ignore the differences between single-particle energies and assume that the strength of the pairing interaction does not depend on the orbitals, $G_{j j'}$= constant for all values of $j$ and $j'$ \cite{kerman}. Richardson, in a series of papers starting with Ref. \cite{Richardson:1966zza}, was able to obtain exact solutions for non-degenerate single particle energies, but with the condition $G_{j j'}$= constant. The limit in which the energy levels are degenerate, but the pairing strength is separable, i.e. 
 $G_{j j'} = |G| c_j^* c_{j'}$, also has exact solutions \cite{Pan:1997rw,Balantekin:2007vs}. In this limit there are  degeneracies between the energies of the $N$ and $N_{\rm max}-N+1$ pair eigenstates, which can be explained by the supersymmetric quantum mechanical structure 
 in the pairing Hamiltonian \cite{Balantekin:2007qr}. Finally the most general solution of the separable pairing Hamiltonian with two non-degenerate orbitals was given in Ref.  \cite{Balantekin:2007ip}. 

\subsection{Analytical solutions with two orbitals}
To illustrate the algebraic approach we consider a specific example, namely the most general separable pairing Hamiltonian with only two orbitals:
\begin{equation}
\label{twoor}
\frac{\hat{H}}{|G|}= \sum_{j} 2\varepsilon_j
\hat{S}_j^0 - \sum_{jj'}c^*_jc_{j'} \hat{S}^+_j
\hat{S}^-_{j'}+\sum_j\varepsilon_j\Omega_j,
\end{equation}
with $\varepsilon_j=\epsilon_j/|G|$. Eigenstates of this Hamiltonian can be written using the step operators  \cite{Balantekin:2007ip}
\begin{equation}
J^+(x)=\sum_j\frac{c_j^*}{2\varepsilon_j-|c_j|^2x}S_j^+
\end{equation}
as
\begin{equation}
\label{state}
J^+(x_1)J^+(x_2)\dots J^+(x_N)|0\rangle . 
\end{equation}
Such a parametrization of the eigenvectors was first used by Bethe in 1931 to find the exact solution of the one-dimensional spin-1/2 
Heisenberg model \cite{bethean} and typically referred to as the Bethe ansatz in the literature.  

Defining
\begin{equation}
\beta=2\frac{\varepsilon_{j_1}-\varepsilon_{j_2}}{|c_{j_1}|^2-|c_{j_2}|^2}
\quad \quad \quad  \delta
=2\frac{\varepsilon_{j_2}|c_{j_1}|^2-\varepsilon_{j_1}|c_{j_2}|^2}
{|c_{j_1}|^2-|c_{j_2}|^2}.
\end{equation}
the energy eigenvalues of the Hamiltonian in Eq. (\ref{twoor}) are 
\begin{equation}
\label{energy}
E_N=-\sum_{n=1}^N\frac{\delta x_n}{\beta-x_n} = \delta N - \sum_{n=1}^N \frac{1}{\beta-x_n}. 
\end{equation}
The parameters $x_k$ in Eqs. (\ref{state}) and  (\ref{energy}) satisfy
the Bethe ansatz equations
\begin{equation}
\sum_{j}\frac{\Omega_j|c_j|^2}{2\varepsilon_j-|c_{j}|^2x_k}
=\frac{\beta}{\beta-x_k} +\sum_{n=1(\neq k)}^N\frac{2}{x_n-x_k}.
\end{equation}
Since the Hamiltonian in Eq. (\ref{twoor}) is Hermitian, the energy eigenvalues in Eq. (\ref{energy}) are real. Hence the solutions of the Bethe ansatz equation, $x_n$, should either be real or the set should include complex conjugate entries. 
Solutions for the limit in which the energy levels are degenerate, but the pairing strength is separable, can also be expressed in terms of solutions of Bethe ansatz equations as shown in Ref. \cite{Pan:1997rw,Balantekin:2007vs}. 

\subsection{Collective Neutrino Oscillations}
In certain astrophysical sites, such as near the neutrinosphere in a core-collapse supernova or neutron-star mergers, density of neutrinos far exceeds the densities of all other particles. In such cases the coherent forward scattering of a test neutrino from other neutrinos dominates neutrino transport. The contribution of such a scattering to the Hamiltonian describing neutrino propagation depends on the angle between momenta of interacting neutrinos. In the single-angle approximation this angle is replaced by an average value. If, for simplicity, one considers only two flavors of neutrinos (e.g., electron and muon neutrino) one can define another SU(2) algebra, the neutrino flavor isospin. For this algebra electron neutrino represents the ``spin-up" and muon neutrino the ``spin-down" components of $J_z$. In the single-angle approximation the Hamiltonian can be written in terms of this SU(2) algebra as \cite{Balantekin:2006tg} 
\begin{equation}
\label{nunu}
H_{\nu\nu} = - \sum_p \frac{\delta m^2}{2p}J^z_p + \eta \sum_{p \neq q} \mathbf{J}_p \cdot \mathbf{J}_q
\end{equation}
where we assumed neutrino mixing, $\delta m^2$ is the difference between squares of masses of the two mass eigenstates that mix, $p$ is the neutrino momentum, and $\eta$ describes the strength of the coherent forward scattering. 
The parameter $\eta$ changes as $\eta \sim 1/{\rm (distance \> from \> the \> neutrino \> source)^3}$. In this equation we denoted the 
neutrino flavor isospin algebra of neutrinos with momenta $\mathbf{p}$ as $\mathbf{J}_p$. Mathematically the neutrino transport Hamiltonian, Eq. (\ref{nunu}), and the pairing Hamiltonian, Eq. (\ref{general}), are very similar. (One should emphasize that they describe very different physics situations. The pairing Hamiltonian, Eq. (\ref{general}) describes a bound-state system whereas the the neutrino 
propagation Hamiltonian describes an adiabatically expanding system). However, in both cases to obtain exact (as opposed to mean-field) solutions one needs to solve appropriate Bethe ansatz equations. 

\section{Solution to Bethe Ansatz Equations}

\subsection{Bethe ansatz}

For both pairing and collective neutrino oscillation problems the most general form of the Bethe ansatz is 
\begin{equation}
\label{bethe}
- \frac{1}{2\mu} - \sum_p^M \frac{j_p}{\omega_p - \lambda_i} = \sum_{j \neq i}^N \frac{1}{\lambda_i - \lambda_j} 
\end{equation}
where $\lambda_i$ are the solutions we seek, $j_p$ is the quantum number associated with the $p$th SU(2) algebra (the Casimir operator of which is $j_p(j_p+1)$), $\omega_p$ are the single-particle energies for the pairing problem or a combination of neutrino masses and energies for the neutrino propagation problem, and $\mu$ is the strength of the two-body term. There are $M$ such SU(2) algebras. One seeks to find $N$-tuple solutions of this equation. Solutions are either real or come in complex-conjugate pairs. For $N=1$ the right-hand side of Eq. (\ref{bethe}) vanishes and the problem of solving  Bethe ansatz equations reduces to the problem of finding real roots of polynomials.  
For $N=2$,  one should find pairs of $\lambda$s the elements of which are either real or complex conjugates of one another. For $N=3$,  
one should find triples of $\lambda$s and so on. For these triplets either all elements are real or one element is real and the other two are complex conjugates of one another. 

Writing down solutions for the eigenstates and eigenvalues of the Hamiltonians in terms of variables to be obtained from solving Bethe ansatz equations simply means such Hamiltonians are exactly solvable and constants of motion are manifest. However to obtain physical quantities you still need to solve the Bethe ansatz equations. These equations are highly nonlinear. Direct numerical attempts may run into stability problems. Electrostatic analogies are useful \cite{Dukelsky:2001dd,Pehlivan:2011hp}, but not sufficient. One needs a numerically stable algorithm.  

\subsection{Solving the Bethe ansatz equations}

Consider the Bethe Ansatz equations associated with two SU(2) algebras with suitably chosen $\omega_1$ and $\omega_2$: 
\begin{equation}
\sum_{k=1(k\neq i)}^{N}\frac{1}{x_{i} -x_{k}}
-\frac{j_2}{x_i}+\frac{j_1}{1-x_i}=0 
\end{equation}
Already in 1914 (before Bethe) Stieltjes showed  \cite{still} that the polynomial 
\begin{equation}
p_N(\lambda) = \prod_{i =1}^{N} (\lambda - x_i)
\end{equation} 
satisfies the hypergeometric equation
\begin{equation}
\lambda(1-\lambda) \frac{d^2p_N}{d \lambda^2} 
+ 2 j_2 \left[ 2j_1-1 \right] \frac{dp_N}{d\lambda} +
N\left(N-2j_1-2j_2-1\right) p_N=0 .
\end{equation}
Hence the roots of the solutions of this equation are the solutions of the Bethe ansatz equation above. It is possible to generalize this observation to the situations with more than two SU(2) algebras. To do so we first define the polynomial: 
\begin{equation}
P(\lambda) = \prod_i (\lambda- x_i) = \exp \left( \sum_i \log (\lambda -x_i) \right)
\end{equation}
and introduce 
\begin{equation}
\Lambda (\lambda) = \frac{dP/d\lambda}{P(\lambda)} = \sum_{i=1}^N \frac{1}{\lambda - x_i} .
\end{equation}
It is easy to show that the quantity $\Lambda (\lambda)$ satisfies the following differential equation: 
\begin{equation}
\label{mastereq}
\Lambda^2 (\lambda) + \frac{d\Lambda}{d\lambda} + \frac{1}{\mu} \Lambda (\lambda)= 2 \sum_p  \frac{j_p}{\lambda - \omega_p} \left[ \Lambda(\lambda) - \Lambda (\omega_p) \right] .
\end{equation}
This equation still presents difficulties, it is a nonlinear differential equation including its own solution in the equation to be solved! The key idea is to calculate $\Lambda (\lambda) $ and its derivatives with respect to $\lambda$ only for $\lambda = \omega_p$. This idea was pursued by many authors in the literature, here we follow the presentation given in  Ref. \cite{Babelon:2007td}. One obtains the following set of equations: 
\begin{equation}
\Lambda^2 (\omega_q) + (1-2 j_q) \Lambda' (\omega_q) + \frac{1}{\mu} \Lambda (\omega_q)= 2 \sum_{p \neq q}  j_p 
\frac{\Lambda(\omega_q ) - \Lambda (\omega_p)}{\omega_q - \omega_p} ,
\end{equation}
where prime denotes derivative with respect to $\lambda$. 
Note that for $j_q=1/2$ the derivative term vanishes and one gets a set of algebraic equations with unknowns $\Lambda (\omega_p) $.  For higher $j_q$ values we can keep taking derivatives. 
For example taking the second derivative of the Eq. (\ref{mastereq}) with respect to  $\lambda$ and then substituting $\lambda = \omega_p$ 
one obtains 
\begin{eqnarray}
2 \Lambda (\omega_q) \Lambda'(\omega_q) &+& (1-j_q)\Lambda''(\omega_q) + \frac{\Lambda'(\omega_q)}{\mu}  \nonumber \\ 
&=& 2 
\sum_{p \neq q} j_p \left[ \frac{\Lambda'(\omega_q)}{\omega_q - \omega_p} + \frac{\Lambda(\omega_q) - \Lambda(\omega_p)}{(\omega_q - \omega_p)^2} \right] \nonumber 
\end{eqnarray}
The second term vanishes for $j_q =1$ and again one gets a set of algebraic equations with unknowns $\Lambda (\omega_p) $ and 
$\Lambda' (\omega_p) $. In principle one can continue this procedure to obtain sets of algebraic equations for all values of $j_q$.

\section{Conclusions}

We discussed a number of pairing-type Hamiltonians the solutions of which can be written in terms of parameters to be obtained from Bethe ansatz equations. These Hamiltonians describe nuclear s-wave pairing, many-neutrino systems in core-collapse supernovae or merging neutron stars, and electrons placed on a number of lattices with spin-spin interactions. In addition we discussed an efficient algorithm to obtain solutions of the Bethe ansatz equations. 

\vskip 0.5cm 
The author would like to thank Taka Otsuka for physics discussions on all aspects of nuclear physics covering three decades. 
This work was supported in part by the US National Science 
Foundation Grant No. PHY-1514695.


\begin{thebibliography}{99}

\bibitem{Dean:2002zx} 
  D.~J.~Dean and M.~Hjorth-Jensen,
  Rev.\ Mod.\ Phys.\  {\bf 75} (2003) 607. 

\bibitem{Otsuka:1978zz} 
  T.~Otsuka, A.~Arima and F.~Iachello,
  Nucl.\ Phys.\ A {\bf 309} (1978) 1. 

\bibitem{Iachello:1979jaq} 
  F.~Iachello, G.~Puddu, O.~Scholten, A.~Arima and T.~Otsuka,
  Phys.\ Lett.\  {\bf 89B} (1979) 1.
  

\bibitem{Balantekin:2000qt} 
  A.~B.~Balantekin and N.~Ozturk,
  Phys.\ Rev.\ D {\bf 62} (2000) 053002.  
  
\bibitem{Balantekin:2006tg} 
  A.~B.~Balantekin and Y.~Pehlivan,
  J.\ Phys.\ G {\bf 34} (2007) 47. 
  
\bibitem{kerman}
A. K. Kerman, Ann. Phys. {\bf 12} (1961) 300.   

\bibitem{Richardson:1966zza} 
  R.~W.~Richardson,
  Phys.\ Rev.\  {\bf 141} (1966) 949. 

\bibitem{Pan:1997rw} 
  F.~Pan, J.~P.~Draayer and W.~E.~Ormand,
  Phys.\ Lett.\ B {\bf 422} (1998) 1. 

\bibitem{Balantekin:2007vs} 
  A.~B.~Balantekin, J.~H.~de Jesus and Y.~Pehlivan,
  Phys.\ Rev.\ C {\bf 75} (2007) 064304. 

\bibitem{Balantekin:2007qr} 
  A.~B.~Balantekin and Y.~Pehlivan,
  J.\ Phys.\ G {\bf 34} (2007) 1783. 

\bibitem{Balantekin:2007ip} 
  A.~B.~Balantekin and Y.~Pehlivan,
  Phys.\ Rev.\ C {\bf 76} (2007) 051001. 
  
 \bibitem{bethean}
H. Bethe, Z. Phys. {\bf 38} (1931) 205.  

\bibitem{Dukelsky:2001dd} 
  J.~Dukelsky, C.~Esebbag and S.~Pittel,
  Phys.\ Rev.\ Lett.\  {\bf 88} (2002) 062501. 

\bibitem{Pehlivan:2011hp} 
  Y.~Pehlivan, A.~B.~Balantekin, T.~Kajino and T.~Yoshida,
  Phys.\ Rev.\ D {\bf 84} (2011) 065008. 

\bibitem{still}
T.J. Stieltjes, 1914 {\em Sur Quelques Theoremes d'Algebre, Oeuvres Completes}, V. 11 (Groningen:Noordhoff). 

\bibitem{Babelon:2007td} 
  O.~Babelon and D.~Talalaev,
  J.\ Stat.\ Mech.\  {\bf 0706} (2007) P06013. 


\end{thebibliography}
\end{document}